\documentclass[aps,11pt,letterpaper,nofootinbib,showpacs,amsmath,amssymb,pdftex11pt]{revtex4}%
\usepackage{latexsym}%
\usepackage[dvips,dvipdfmx,hypertex]{hyperref}%

\def\cF{{\mathcal F}}
\def\cH{{\mathcal H}}

\def\cW{{\mathcal W}}

\DeclareMathAlphabet{\mathpzc}{OT1}{pzc}{m}{it}

\newcommand{\beq}{\begin{equation}}
\newcommand{\beqn}{\begin{equation}\nonumber}
\newcommand{\eeq}{\end{equation}}
\newcommand{\bea}{\begin{eqnarray}}
\newcommand{\bean}{\begin{eqnarray}\nonumber}
\newcommand{\eea}{\end{eqnarray}}

\begin{document}

\title{Quantum gravitational dust collapse does not result in a black hole}
\author{Cenalo Vaz}
\address{Department of Physics, University of Cincinnati, Cincinnati, OH 45221-0011.}
\email{\tt Cenalo.Vaz@uc.edu}

\begin{abstract}
Quantum gravity suggests that the paradox recently put forward by Almheiri et. al. (AMPS) can be
resolved if matter does not undergo continuous collapse to a singularity but condenses on the 
apparent horizon. One can then expect a quasi-static object to form even after the gravitational 
field has overcome any degeneracy pressure of the matter fields. We consider dust collapse. If 
the collapse terminates on the apparent horizon, the Misner-Sharp mass function of the dust ball 
is predicted and we construct static solutions with no tangential pressure that would represent 
such a compact object. The collapse wave functions indicate that there will be processes by which 
energy extraction from the center occurs. These leave behind a negative point mass at the center 
which contributes to the total energy of the system but has no effect on the the energy density 
of the dust ball. The solutions describe a compact object whose boundary lies outside its 
Schwarzschild radius and which is hardly distinguishable from a neutron star.

\end{abstract}

\pacs{04.60.-m, 04.70.Dy, 97.60.Lf, 97.60.Jd}
\maketitle

\section{Introduction}

The final fate of gravitational collapse has long been a mystery. Classical collapse models suggest 
that a star that is massive enough to overcome all degeneracy pressures will undergo collapse beyond 
the apparent horizon \cite{genc,hael73} eventually forming a naked or covered singularity of spacetime, 
depending on the initial conditions. But there is something deeply unsatisfying about this picture 
since it does not take into account quantum gravity, which is expected to play a significant role in 
the final stages of collapse.

If the collapse begins with initial data that lead to the formation of a naked singularity \cite{psj} 
then a semi-classical treatment of the radiation (assuming the validity of effective field theory) 
from the singularity suggests that the final stages will be catastrophic \cite{vwns,bsvw}. It is not 
known what the final fate of such a collapse is: either the collapsing star will dissipate entirely or a 
remnant will attempt to form a covered  singularity. However, if the initial data are such as to lead to 
the formation of a covered singularity and an event horizon forms then, as Hawking pointed out \cite{haw75},
the semi-classical theory would yield thermal radiation from the point of view of the observer who 
remains outside the black hole provided that the freely falling observer detects nothing unusual while 
crossing the horizon. The semi-classical analysis would seem to suggest that information is lost if the 
black hole evaporates completely, since what is left is a density matrix and not a wave function. But if 
the quantum theory is unitary then either (a) the evaporation is not in fact thermal and the Hawking radiation 
is pure or (b) the thermal evaporation process should, by an as yet unknown mechanism, leave behind a stable 
remnant that contains all the information that fell into the hole. The second option is difficult to imagine 
since a relatively small object would be required to possess a huge degeneracy while remaining stable. 
Moreover, it is ruled out if we assume that quantum gravity is CPT invariant. 

This leaves just the first option, that the Hawking radiation is pure. In 1993, Susskind {\it et. al} 
\cite{stu93}, building on the work of 't Hooft \cite{thof85} and Preskill \cite{pres92}, proposed that 
the unitarity of the Hawking radiation could be preserved if information is both emitted at the horizon and 
passes through it, so an observer outside would see it in the Hawking radiation and an observer who falls 
into the black hole would see it inside but no single observer would be able to confirm both pictures. 
Although there is no precise mechanism by which it is can be said to occur, thought experiments that appear 
to support this picture of ``Black Hole Complementarity'' rely on three fundamental assumptions, {\it viz.,} 
(a) the unitarity of the Hawking radiation, (b) the validity of effective field theory outside a ``stretched'' 
horizon and (c) the equivalence principle. Recently, however, Almheiri et. al. (AMPS) have argued that the 
three assumptions are logically inconsistent and would lead to a violation of the strong subadditivity of 
the entanglement entropy \cite{amps12,bpz09,dyhz11}. To resolve the paradox the authors suggested giving up 
the third assumption, {\it i.e.,} the equivalence principle. 

But Hawking has proposed an intriguing alternative, suggesting that no event horizon would form in the first 
place if somehow the collapse did not continue beyond the apparent horizon \cite{haw14}. If indeed no event horizon 
is formed, the entire discussion about information loss becomes moot. Yet, one is left with the 
question of how the system evolves after the formation of the apparent horizon. 
There appears to be ample experimental evidence supporting the existence of very massive, quasi-stable, compact 
objects located in galactic centers that are consistent with black holes, although it is not known for certain if 
these supermassive configurations are indeed black holes with event horizons. In this paper, we will examine 
Hawking's conjecture as it relates to dust collapse, by re-examining some results of an exact quantization 
\cite{vw01,genqc} of the LeMa\^\i tre-Tolman-Bondi family of solutions \cite{ltb}. 

In previous work  \cite{vwvl} we have shown that two kinds of functional solutions (analogous to plane waves) of the 
Wheeler-DeWitt equation for dust collapse may be given. In one, dust shells coalesce onto the apparent 
horizon on both sides of it. Exterior, infalling waves representing the collapsing shells of dust are accompanied 
by interior, outgoing waves, which are produced with a relative probability given by the Boltzmann factor at the 
Hawking temperature of the shells. These interior waves, which are of quantum origin, represent an interior Unruh 
radiation. In the other solution, dust moves away from the apparent horizon on both sides of it. Interior, 
infalling waves representing the continued collapse of the dust shells across the apparent horizon are accompanied 
by exterior, outgoing waves, which are produced with a relative probability again given by the Boltzmann factor. 
at the Hawking temperature. These latter ougoing waves represent the exterior Unruh radiation. 

Continued collapse across the apparent horizon from an initial diffuse state and to a central singularity can be achieved 
by combining the two solutions and requiring the net flux to vanish at the apparent horizon as in \cite{vwvl}. The net effect 
is that the collapse is accompanied by Unruh radiation in the exterior, as is well known \cite{hrcond}, but ends in a 
central singularity. However, if the collapse does not continue past the apparent horizon, there will be no exterior 
radiation during the collapse. Furthermore, as the shells coalesce on the apparent horizon, no event horizon will form 
and the AMPS paradox is resolved. This picture is captured by the first of the exact solutions of the Wheeler-DeWitt 
equation discussed in the previous paragraph \cite{vaz14}. 

In this paper, we will examine the consequences of taking seriously the possibility that continued collapse does not occur, 
{\it i.e.,} that {\it quantum} collapse is described by the first solution described above. The collapsing matter is then 
accompanied by Unruh radiation in the interior of the apparent horizon. In this case, we expect to end up with a spherically 
symmetric, quasi-static configuration of finite extension and with a specific mass function as the end 
state of the collapse. Even though no {\it classical}, static, extended dust configuration can exist, we will show 
that the interior Unruh radiation that accompanies the infalling dust shells during the collapse will generate the 
conditions appropriate for a quasi-static configuration to exist. In effect it creates a negative mass point source at 
the center of the star, which is enveloped by the collapsed matter. We allow for radial but no tangential pressure. This 
is in keeping with the midi-superspace quantization that informs our construction \cite{vw01,genqc}. With the inclusion of 
a constant negative vacuum energy and radial pressure, unique static solutions exist. There are no horizons and the matter 
itself extends to twice the Schwarzschild radius. 

In section II we briefly summarize our previous work on the wave functionals describing the collapse. In section III we 
construct the static, spherically symmetric solutions described above and analyze the solutions. In section IV we 
estimate the size of the central negative mass. We conclude with a discussion on the our results and possible implications 
for future observations in section IV. We take $\hbar = c = 1$ in what follows.

\section{Quantum Dust Collapse}

Dust collapse in any dimension, with or without a cosmological constant, is described by the LeMa\^\i tre-Tolman-Bondi 
family of solutions \cite{ltb}. In comoving and synchronous coordinates, $(t,\rho,\theta,\phi)$, one has
\beq
ds^2 = d\tau^2 - \frac{R'(\tau,\rho)^2}{1+f(\rho)} d\rho^2 - R^2(\tau,\rho) d\Omega^2,
\label{ltb}
\eeq
where the area radius, $R(\tau,\rho)$ obeys the Einstein equation
\beq
\dot R(\tau,\rho) = \sqrt{f(\rho) + \frac{2G F(\rho)}{R(\tau,\rho)} + \frac 13 \Lambda R^2(\tau,\rho)}
\label{ltbeom}
\eeq
and the energy density is given by
\beq
\epsilon(\tau,\rho) = \frac{F'(\rho)}{R^2(\tau,\rho) R'(\tau,\rho)}.
\eeq
$\Lambda$ is the cosmological constant. There are two integration functions, $F(\rho)$ and $f(\rho)$, that are 
interpreted as the twice the gravitational (Misner-Sharp) mass contained within a shell located at $\rho$ and the 
total energy contained within the same shell respectively. They are the ``mass'' and ``energy'' functions 
of the collapse \cite{psj,plkr}.

By considering the expansion of an outgoing, radial null congruence, 
\beq
\Theta = \frac{2R'(\tau,\rho)}{R(t,\rho)}\left[1-\sqrt{\frac{2G F(\rho)}{R(\tau,\rho)} + \frac 13 \Lambda 
R^2(\tau,\rho)}\right],
\eeq
one sees that the condition for trapping is met when 
\beq
\frac{2GF(\rho)}{R(\tau,\rho)} + \frac 13 \Lambda R^2(\tau,\rho) = 1,
\eeq
which can be used to determine the time of formation, $\tau_\text{ah}(\rho)$, of the apparent horizon once
a solution of \eqref{ltbeom} is determined. 

The canonical dynamics of the collapsing dust shells is described by embedding the spherically symmetric ADM metric in the 
LTB spacetime of \eqref{ltb}. After a series of canonical transformations \cite{vw01,ku94,kubr95}, they are described in 
a phase space consisting of the dust proper time, $\tau(t,r)$, the area radius, $R(t,r)$, the mass density, $\Gamma(r) = 
F'(r)$, and their conjugate momenta, $P_\tau(t,r)$, $P_R(t,r)$ and $P_\Gamma(t,r)$ respectively by two constraints,
\bea
\cH_r &=& \tau' P_\tau + R' P_R - \Gamma P_\Gamma' \approx 0\cr\cr
\cH &=& P_\tau^2 + \cF P_R^2 - \frac{\Gamma^2}{\cF} \approx 0,
\label{constraints}
\eea
where the prime denotes a derivative with respect to the ADM radial label coordinate, $r$, and 
\beqn
\cF~ \stackrel{\text{def}}{=}~ 1 - \frac{2GF}R - \frac 13 \Lambda R^2.
\eeq
The apparent horizon occurs when $\cF = 0$. In the absence of a cosmological constant, this says that on the apparent 
horizon the physical radius of each shell is given by 
\beq
R(\tau_\text{ah},\rho) = 2GF(\rho).
\label{massfn}
\eeq

Dirac's quantization procedure may be employed to turn the classical constraints in \eqref{constraints} into quantum 
constraints, which act on wave functionals. The Hamiltonian constraint then yields a formal Wheeler-DeWitt equation and the 
momentum constraint imposes diffeomorphism invariance. We begin with an ansatz for the wave functional \cite{vw01},
\beq
\Psi[\tau,R,\Gamma] = \exp\left[-\frac i2\int dr \Gamma(r) \cW(\tau(r),R(r),\Gamma(r))\right],
\label{ansatz}
\eeq
which automatically satisfies the momentum constraint if $\cW$ has no explicit dependence on $r$. The Wheeler-DeWitt equation 
must be regularized before solutions can be obtained. This regularization was performed on a one dimensional lattice 
\cite{vsw04,kmv06} given by a discrete set of points, $r_i$, representing dust shells and separated by a spacing $\sigma$. 
One then finds that the ansatz in \eqref{ansatz} yields a product of what may be thought of as shell wave functions,
\begin{widetext}
\beq
\Psi = \lim_{\sigma\rightarrow 0}\prod_i \psi_i (\tau_i,R_i,\Gamma_i) = \lim_{\sigma\rightarrow 0}\prod_i e^{\omega_i b_i} 
\times \exp\left\{-i\omega_i \left[a_i \tau_i \pm \int^{R_i} dR_i \frac{\sqrt{1-a_i^2 \cF_i}}{\cF_i}\right]\right\},
\label{lattice}
\eeq
\end{widetext}
with a well defined continuum limit ($\sigma\rightarrow 0$), where $a_i = 1/\sqrt{1+f_i}$ is related to the energy function 
and $\omega_i = \sigma \Gamma_i/2$. Diffeomorphism invariance also requires that both $a_i$ and $b_i$ depend on $r$ via the 
mass function, {\it i.e.,} $a_i = a_i (F_i)$ and $b_i = b_i(F_i)$.

These solutions are defined everywhere except at the apparent horizon. Thus there are ``exterior'' wave functions that 
must be matched to ``interior'' wave functions across the horizon. As can be seen, however, the phases of the interior 
and exterior wave functions diverge there. A standard technique used in such cases is to analytically continue the solutions 
into the complex plane. This technique was used to derive the Hawking radiation as a tunneling process in \cite{parwil00}. 
Thus, analytically continuing into the complex $R$  plane, taking $\cF_i = \epsilon \exp[i\varphi]$, with $\epsilon>0$, and 
comparing them at $\varphi = \pi/2$. One then finds two sets of matched solutions, with support everywhere; the first is 
given by \cite{vwvl}
\begin{widetext}
\beq
\psi_{i,\text{col}}^{(1)}(\tau_i,R_i,F_i) = \left\{\begin{matrix}
e^{\omega_ib_i} \times \exp\left\{-i\omega_i \left[a_i \tau_i + \int^{R_i} dR_i
\frac{\sqrt{1-a_i^2\cF_i}}{\cF_i}\right]\right\} & \cF_i>0\cr
e^{-\frac{\pi\omega_i}{g_{i,h}}}\times e^{\omega_ib_i} \times \exp\left\{-i\omega_i
\left[a_i \tau_i + \int^{R_i} dR_i \frac{\sqrt{1-a_i^2\cF_i}}{\cF_i}\right]\right\} & \cF_i < 0
\end{matrix}\right.
\label{wfn1}
\eeq 
\end{widetext}
and the second by
\begin{widetext}
\beq
\psi_{i,\text{col}}^{(2)}(\tau_i,R_i,F_i) = \left\{\begin{matrix}
e^{-\frac{\pi\omega_i}{g_{i,h}}}\times e^{\omega_ib_i} \times \exp\left\{-i\omega_i
\left[a_i \tau_i - \int^{R_i} dR_i \frac{\sqrt{1-a_i^2\cF_i}}{\cF_i}\right]\right\} & \cF_i > 0 \cr
e^{\omega_ib_i} \times \exp\left\{-i\omega_i \left[a_i \tau_i - \int^{R_i} dR_i
\frac{\sqrt{1-a_i^2\cF_i}}{\cF_i}\right]\right\} & \cF_i < 0,
\end{matrix}\right.
\label{wfn2}
\eeq
\end{widetext}
where $g_{i,h}=\partial_R\cF(R)|_{R_{i,h}}/2$ is the surface gravity of the $i^\text{th}$ shell at the apparent horizon.

These are the shell wave functions we described in the introduction. The first (in \eqref{wfn1}) represents a flow 
toward the apparent horizon on both sides of it: an infalling shell in the exterior is accompanied by an interior, 
outgoing shell, produced with a relative probability determined by the Boltzmann factor at the Hawking temperature of 
the shell. The second (in \eqref{wfn2}) describes a flow away from the apparent horizon: an infalling shell in the interior, 
which represents its continued collapse past the apparent horizon and to a central singularity is accompanied by an exterior, 
outgoing shell, with a relative probability also determined by the Boltzmann factor. It represents the thermal radiation 
in the exterior. 

One might in principle be interested in constructing wave packets that represent an evolution from a configuration in 
which the dust cloud begins far from the apparent horizon. Such a wave packet would serve to clarify the semi-classical 
description of the collapsing ball and would be constructed by superposing the solutions given above with different 
energies, $a_i(F_i)$. This difficult problem, which is currently under investigation, does not seem feasable at present as 
both the factor ordering and the diffeomorphism invariance depend on the energy function \cite{kmv06}. Nevertheless, some useful 
conclusions can be drawn from the ``plane wave'' solutions we have presented above, as one does, for example, in ordinary 
quantum scattering theory. We notice that if we take the wave functions in \eqref{wfn1} to form the basis for the {\it quantum} 
collapse of dust then there will be thermal Unruh radiation inside the apparent horizon but no thermal radiation outside, 
accompanying the collapse. There will also be no continued collapse to a central singularity; the collapse would terminate at 
the apparent horizon ($\cF_i=0$), which agrees with Hawking's proposal \cite{haw14,vaz14}. These conclusions would hold true 
even if one could find a way to construct diffeomorphism invariant wave packets from \eqref{wfn1} representing the collapse. 

\section{A Quasi-Classical Configuration}

As there is good experimental evidence for the existence of very masive, quasi-stable compact objects, we look for static, 
spherically symmetric solutions of Einstein's equations satisfying the following criteria:
\begin{itemize}
\item the collapsed dust ball should occupy a finite region and posesses an energy density that is characteristic of a dust 
cloud that has condensed onto its apparent horizon, {\it i.e.,} given by \eqref{massfn},
\item the solutions should incorporate the effect of the internal Unruh radiation that has occurred during the collapse phase and
\item they must match smoothly to the Schwarzschild vacuum at the boundary.
\end{itemize}
Within the dust ball the metric will be of the form
\beq
ds^2 = e^{2A} dt^2 - e^{2B} dr^2 - r^2 d\Omega^2,
\eeq
where $A=A(r)$, $B=B(r)$ and $r$ represents the physical radius. In this coordinate system, if we take the components 
of the stess-energy to be ${T^\mu}_\nu = \text{diag}(-\varepsilon(r),p_r(r), p_\theta(r),p_\theta(r))$ but impose no 
equations of state, the field equations are
\bea
&&1 - e^{2B} -2rB' = -8\pi G r^2e^{2B}\varepsilon\cr\cr
&& 1 - e^{2B} + 2 r A' = 8\pi G r^2 e^{2B}p_r\cr\cr
&& rA'^2 -B'+A'(1-rB')+rA'' = 8\pi G r e^{2B} p_\theta,\cr
&&
\label{ein}
\eea
where a prime indicates a derivative with respect to the radius, $r$. The conservation of energy-momentum gives a 
constraint,
\beq
\varepsilon A' + p_r' + p_r\left[\frac 2r + A'\right] - \frac 2r p_\theta = 0,
\label{cons}
\eeq
which represents the condition for static equilibrium. Two of the stress-energy components may be chosen arbitrarily 
and then the third is determined by either Einstein's equations or by the conservation law. Below we will choose the 
energy density and set the tangential pressure to zero.

The first equation in \eqref{ein} may be re-expressed as 
\beq
[r(1-e^{-2B})]' = 8\pi G r^2 \varepsilon,
\label{gammaeqn}
\eeq
which is straightforwardly integrated to give
\beq
r(1-e^{-2B}) = 8\pi G\int^r dr' r'^2 \varepsilon(r') - r_0,
\label{int1}
\eeq
where $r_0$ is an integration constant. This is usually set to zero in stellar models to avoid a central singularity, but 
we will not do so here for reasons that will become clear in the following. The Misner-Sharp mass function of the dust
is to be identified with the integral on the right,
\beq
F(r) = 4\pi\int^r dr' r'^2 \varepsilon(r).
\eeq
Now, according to \eqref{massfn}, the mass function that may be expected of a dust ball whose collapse has 
terminated at the apparent horizon is 
\beq
F(r) = \frac r{2G},
\label{massfncl}
\eeq
for a total gravitational mass of $M_{ms}=F(r_b)=r_b/2G$, where $r_b$ denotes its boundary. It 
corresponds to an energy density of 
\beq
\varepsilon(r) = \frac 1{8\pi G r^2}
\label{enden}
\eeq
and \eqref{int1} gives
\beq
e^{2B} = r/r_0.
\label{Bfn}
\eeq
We see that the constant $r_0>0$ is essential and cannot be discarded. Without it there do not exist 
real solutions for $B(r)$ with the desired mass function, even if pressure is included. Strictly it 
describes a negative mass point source the center. Such a negative mass source is actually predicted 
by the wave functions in \eqref{wfn1} to form {\it during} the collapse as energy is extracted from 
the center by the interior, outgoing Unruh radiation that accompanies the exterior, collapsing shells. 
This  process of energy extraction from the center continues until the collapse terminates. In the 
next section we will estimate its size.

With $B(r)$ given in \eqref{Bfn} and no tangential pressure, we solve the Riccati equation in \eqref{ein} 
for $A(r)$ and find
\beq
ds^2 = r^2\left(1+\frac\gamma{r^{3/2}}\right)^2 dt^2 - \frac r{r_0} dr^2 - r^2 d\Omega^2,
\eeq
where $\gamma$ is another integration constant. 
There are curvature singularities at $r=0$ and at $r=(-\gamma)^{2/3}$.  To avoid the singularity at 
$r=(-\gamma)^{2/3}$, either $\gamma$ must be positive or $(-\gamma)^{2/3}$ must lie outside the outer 
boundary of the collapsed star, where the solution no longer applies.  We will soon show that the second 
condition cannot be met. 

We determine the pressure directly from the second equation in \eqref{ein}
\beq
p_r(r) = -\frac 1{8\pi G r^2}\left[1 - \frac{3r_0}r\left(1+\frac \gamma{r^{3/2}}\right)^{-1}\right],
\eeq
so with $\gamma\geq 0$ our solutions are well behaved except at the singular center and they obey 
the weak energy conditions.

If $r_b$ denotes the outer boundary of the collapsed star, we want to match the interior geometry to an 
external vacuum, described by the Schwarzschild metric
\beq
ds^2 = f(R) dT^2 - f^{-1}(R) dR^2 - R^2 d\Omega^2,
\eeq
where 
\beqn
f(R) = \left(1-\frac{2GM_s}R\right)
\eeq
and $M_s$ is the Schwarzschild mass of the dust ball. The junction conditions require that 
\bea
&&R_b = r_b, ~~ T_b = \frac{e^{A(r_b)}}{\sqrt{f(r_b)}}~ t\cr\cr
&&e^{-B(r_b)} = \sqrt{f(R_b)},~~ 2 A'(r_b) = (\ln f)'|_{R_b}
\eea
and therefore
\beq
r_0 = r_b-r_s,~~ \gamma = 2 r_b^{3/2}\left(1 - \frac{3r_s}{2r_b}\right),
\label{sff}
\eeq
where we have let $r_s = 2GM_s$ be the Schwarzschild radius. 

The first condition says that the physical radius of the boundary must lie outside its Schwarzschild radius. 
Therefore, as expected, the Schwarzschild mass of the star is less than the Misner-Sharp mass of the dust,
\beq
M_{s} = \frac{r_s}{2G} = \frac{r_b-r_0}{2G} = M_{ms}- M_0,
\eeq
by precisely the negative central mass, $-M_0$. If $\gamma \geq 0$, the second condition requires that $r_b 
\geq 3r_s/2$. But, for $\gamma < 0$ a regular solution is obtained only if we require the singularity to 
lie outside the boundary of the star. According to \eqref{sff}, this can happen if 
\beq
2\left(\frac{3r_s}{2r_b} -1\right) > 1,
\eeq
but this would imply that $r_s > r_b$. As this is not possible, the star will be singularity free (except at 
the center) only if $r_b \geq 3r_s/2$. This implies that $r_b \leq 3r_0$ and $r_s \leq 2r_0$.

\section{Estimating $r_0$}

We can provide a simple estimate of the radius, $r_0$, as follows. The energy extraction from the center 
occurs during the collapse because every 
collapsing shell is accompanied by an interior, outgoing wave, which will extract energy from the center. 
We want to estimate how much energy is extracted in this process. For the given mass function, the energy 
density of the dust is constant, $\Gamma = 1/2G$. If $\sigma$ represents the shell thickness, the average 
energy, $\omega_i$, of each shell will also be constant, $\omega_i = \omega = \sigma\Gamma/2 = \sigma/4G$. 

The collapse of the $i^\text{th}$ shell will have been accompanied by the emission in the interior 
of an outgoing wave of the same frequency, with a probability that is given by the Boltzmann factor, $e^{-\beta_i 
\omega}$, at the Hawking temperature, $\beta_i = 2\pi r_i$, of the shell. It follows that the average energy 
of the outgoing shell is $\langle \omega\rangle = \omega e^{-\beta_i\omega}$ and, to get the total energy 
extracted, we must sum over all collapsed shells,
\beq
M_0 = \frac 1\sigma\int_0^{r_b} dr \omega e^{-2\pi \omega r} = \frac 1{2\pi\sigma}\left[1-e^{-2\pi\omega r_b}
\right].
\eeq
Replacing $\omega$ by $\sigma/4G$ and taking the limit as the shell spacing approaches zero then gives
\beq
M_0 = \frac{r_b}{4G} = \frac 12 M_{ms},
\eeq
which implies that $r_0 = r_b/2$. By the matching conditions, it follows that $r_0 = r_s$, therefore the 
region of negative energy occupies the Schwarzschild radius of the star. Although it extends to half 
the boundary radius of the collapsed dust ball and is necessarily surrounded by a cloud of ordinary 
matter, this is a surprisingly large length scale over which quantum gravitational effects should predominate. 
There is no event horizon. A photon, emitted near the boundary of this cloud, would experience a relatively 
tame redshift of 
\beq
z = \sqrt{\frac{r_b}{r_0}} - 1 = \sqrt{2}-1 \approx 0.414,
\eeq
which is compatible with the gravitational redshift of neutron stars of low core densities \cite{lind84}, 
suggesting that, in a collapse of realistic matter, quantum gravity could ``kick in'' much before previously 
imagined, very near the time at which nuclear densities are achieved. This is consistent with the idea that 
in extreme conditions quantum gravity may be relevant on distance scales much larger than previously anticipated.

\section{Discussion}

In this paper we have speculated on the consequences of a simple quantum model of dust collape. We have argued 
that the AMPS paradox is avoided if continued collapse does not occur and all dust shells coalesce onto
the apparent horizon. We showed that the collapse process is then accompanied by Unruh radiation within the apparent 
horizon. We argued that the effect of the interior Unruh radiation is energy extraction from the center of the 
cloud, leaving behind a negative mass singularity as the cloud settles into a quasi-stable equilibrium. 

Stable classical solutions, with the given mass function and including pressure were determined. The solutions 
are governed by two parameters, the Schwarzschild radius, $r_s$, of the dust ball, equivalently its mass as 
measured by a distant observer, and the boundary radius, $r_b$. The difference between the two is the radius, 
$r_0$, of a region inside of which the total energy is negative. 
There are strong constraints on the parameters $r_0$, $r_s$ and $r_b$ if the interior geometry is required to be 
well behaved everywhere (except at the center). We have shown that the $r_0$ should extend to more than 
one half the Schwarzschild radius and more than one third the radius of the entire star, so it will occupy 
a significant fraction of the star. A more detailed analysis of the Unruh radiation from the center during the 
collapse indicated that $r_0 = r_s = r_b/2$.

The effective energy momentum tensor which sources the Einstein equations in \eqref{ein} is presumed to contain 
quantum gravitational corrections incorporating the back reaction of the Unruh radiation. Beginning, to the 
best of our knowledge, with \cite{fv81}, in which it was argued that quantum gravitational corrections can make 
gravity repulsive at very high densities, many attempts have been made at modeling the radiation back reaction 
via an effective action for the gravitational field, but this approach has proved to be challenging and remains 
poorly understood. More recently, an interesting model of homogeneous, perfect fluid collapse was studied in 
\cite{bmm14}, where it was argued that the repulsive nature of the quantum corrections at short distances would 
cause the collapse to bounce. A general action modeling dimensionally reduced, spherical gravity with a radiation 
term taken from the two dimensional conformal anomaly was examined in \cite{tk14}, but collapse was not discussed 
there. In a different approach, the Unruh radiation was included explicitly in a numerical study of dust collapse 
in \cite{mhp14}. The authors concluded that the collapse results in a bounce prior to crossing the apparent horizon 
and found, in the cases addressed, that the effective mass of the star is reduced through Unruh radiation by 
a factor of two, while the star shrinks preserving $r_b/r_s \gtrsim 2$. The latter conclusions are strikingly 
similar to our own findings, although in this paper we have asked for static solutions and there is 
a difference between the predictions of our wave functional \eqref{wfn1} and the model of \cite{mhp14} in that 
the Unruh radiation in our model is confined to the interior of the apparent horizon. It therefore remains to show 
how the object we have described and in particular the negative energy central singularity, which weakens 
gravity, may form as the end state of the collapse of matter obeying reasonable energy conditions with regular 
initial data. This is necessary for the solution proposed above to constitute a resolution of the information 
paradox and we hope to report on it in the future.

Even in this simple dust model the picture that emerges is somewhat different from the traditional view of a black 
hole.  The collapsed dust object occupies twice its Schwarzschild radius, there is no horizon and the spacetime 
geometry is regular everywhere except at the center. What is being encountered has more in common with other Compact 
Stars (CSs) such as neutron stars, except that what holds the system up is not the matter equation of state (EOS) 
but vacuum energy. The traditional radius of the hole (the Schwarzschild radius) is in fact surrounded by a matter 
cloud. Outgoing radiation from close to boundary of this cloud should not suffer a gravitational redshift much larger than 
from very compact neutron stars. Observationally, however, assuming that the general conclusions of our model continue 
to hold in the presence of more realistic matter, there will be some differences.

An obvious difference between the two types of compact objects is that while the mass of an ordinary CS is limited 
by the matter EOS, there is no limit to how massive a quantum black hole may be. If the matter EOS is assumed to 
be solely responsible for holding up the star, then the recently discovered CSs of mass $\sim$ $2 M_\odot$ 
\cite{dph10,afw13} appear to rule out exotic matter EOSs, which tend to become soft at high densities. Under the same 
assumption, this leaves ``ordinary'' (nucleonic) matter EOSs with comparatively large radii $> 11$ km for a $2 M_\odot$ 
CS \cite{dph10}. Quantum black holes would be both more massive and possess smaller radii than neutron stars, but 
larger radii than classical black holes of the same mass. Therefore it is necessary to measure the radius of CSs in 
a precise and model independent way to provide this information. While this has proved difficult so far, the proposed 
Large Observatory for X-ray timing (LOFT) has claimed to be able to measure the radii of some CSs with a precision of 
up to 1 km \cite{fer12}. 

Another difference between them will be their luminosities in the presence of accretion flows such as would occur in 
X-ray binaries or in galactic centers where supermassive black holes are thought to exist. One may expect accretion 
onto the surface of an ordinary neutron star to lead to higher luminosities than accretion onto the surface of a 
quantum black hole because an accreting shell of matter encounters a hard surface as it collapses onto an ordinary 
neutron star, but the quantum theory dictates that it should slow down and coalesce onto the apparent horizon as it approaches 
the ``surface'' of a quantum black hole. Accreting quantum black holes will therefore look fainter than accreting neutron 
stars.  The reason for the darkness of the quantum black hole is quantum mechanics and not the absence of a surface, but 
the outcome agrees qualitatively with the predictions of \cite{nacl08,bln09}.

Very large compact objects, such as the supermassive black holes that are thought to exist at the centers of galaxies 
make excellent candidates for verifying or falsifying the existence of quantum black holes, if their radius can be 
determined accurately. In the near future, observations of the supermassive black hole Sgr A* by the Event Horizon 
Telescope (EHT) are expected to be sensitive to distance scales of better than a horizon length in the 
1 mm range and direct measuremnts of Sgr A*'s size are expected to become possible  \cite{hetal14,ketal13}. 

Finally, we also mention that a recent study of the periodic modulation in the intensity vs. frequency spectrum 
of galactic centers seems to support the {\it similarity} between behaviors of certain pulsars and supermassive black 
holes \cite{ebor13}. These issues are under investigation, as is also the problem of constructing wave packets 
representing a collapsing dust ball.
\bigskip

\noindent{\bf Acknowledgement}
\medskip

\noindent I thank L.C.R. Wijewardhana for useful discussions.

\end{document}